\documentclass[conference,twocolumn]{IEEEtran}

\usepackage{amsmath,epsfig,bm,amssymb,amsthm}
\usepackage{psfrag}
\usepackage{cite}

\ifCLASSINFOpdf
\else
\fi

\usepackage{amsmath,epsfig,bm,amssymb,amsthm,balance}

\newcommand{\dd}{{\mathrm d}}
\newcommand{\sd}{{\mathrm {sd}}}

\begin{document}
%
\title{Differential Amplify-and-Forward Relaying in Time-Varying Rayleigh Fading Channels
\thanks{The authors are with the Department of Electrical and Computer Engineering,
University of Saskatchewan, Saskatoon, Canada, S7N5A9.
Email: m.avendi@usask.ca, ha.nguyen@usask.ca.}}

\author{\IEEEauthorblockN{M. R. Avendi and Ha H. Nguyen}
\IEEEauthorblockA{Department of Electrical and Computer Engineering\\
University of Saskatchewan\\
Saskatoon, Canada, S7N5A9\\
Email: m.avendi@usask.ca, ha.nguyen@usask.ca}
}

\maketitle

\begin{abstract}
\label{abs}
This paper considers the performance of differential amplify-and-forward (D-AF) relaying over time-varying Rayleigh fading channels. Using the auto-regressive time-series model to characterize the time-varying nature of the wireless channels, new weights for the maximum ratio combining (MRC) of the received signals at the destination are proposed. Expression for the pair-wise error probability (PEP) is provided and used to obtain an approximation of the total average bit error probability (BEP).
The obtained BEP approximation clearly shows how the system performance depends on the auto-correlation of the direct and the cascaded channels and an irreducible error floor exists at high signal-to-noise ratio (SNR). Simulation results also demonstrate that, for fast-fading channels, the new MRC weights lead to a better performance when compared to the classical combining scheme. Our analysis is verified with simulation results in different fading scenarios.
\end{abstract}




\IEEEpeerreviewmaketitle


\section{Introduction}
\label{se:intro}
The employment of other wireless users as relays in a communication network has been shown to be a promising technique to improve the performance of wireless links \cite{coop-laneman}. This technique is especially important for mobile applications, which are not able to support multiple transmit antennas due to insufficient space at the mobile units. In a relay network, the transmission process is divided into two phases. In phase I, the source transmits data to the relays. In phase II, relays perform a decode-and-forward (DF) or amplify-and-forward (AF) strategy to send the received data to the destination \cite{coop-laneman}.

In an AF network, depending on the type of modulation, the relay and the destination may need the channel state information (CSI) of both channels. To avoid channel estimation at the relay and the destination, differential AF (D-AF) scheme has been considered in \cite{DAF-Liu,DAF-DDF-QZ,General-DAF} which only needs the second-order statistics of the channels at the relay and the destination. In the absence of CSI, a set of fixed weights, based on the second-order statistics, were used to combine the received signals from the relay-destination and the source-destination links. However, all the previous works assume a slow-fading situation and show that the performance of D-AF is about 3-4 dB worse than its coherent version. For future reference, the differential detection scheme developed for slow-fading scenario shall be refereed to as ``classical differential detection'' (CDD). In practice, the everyday increasing of the mobility of users leads to time-selective channels and it is therefore important to consider performance of relaying systems under practical channel variation scenarios.

This paper studies the performance of D-AF in \emph{time-varying} Rayleigh fading channels. The detection scheme developed for such time-varying channels shall be referred to as ``time-varying differential detection'' (TVD). The channels from the source to the relay and the destination and from the relay to the destination change continuously according to the Jakes model \cite{microwave-jake}. The direct channel is modelled with a first order auto-regressive model, AR(1) \cite{AR1-ch}. Also, based on the AR(1) model of individual Rayleigh-faded channels, a time-series model is proposed to characterize the time-varying nature of the cascaded channel. Considering the channel variations, new fixed weights for combining the received signals at the destination are given. Since, analyzing the performance of the system using the fixed combining gains is too complicated (if not impossible), the performance of the system using optimum combining weights is analyzed and the result is used as a lower bound for the system error performance. The pair-wise error probability (PEP) is obtained and used to approximate the average bit error probability (BEP) using nearest neighbour approximation. It is shown that an error floor exits at high signal-to-noise ratio (SNR) region, which is related to the auto-correlations of both the cascaded channel and the direct channel. Simulation results reveal that the TVD with the proposed weights always outperforms the CDD in time-selective channels.

The outline of the paper is as follows. Section \ref{sec:system} describes the system model. In Section III the channel model and the differential detection of D-AF relaying with MRC technique over time-varying channels is developed. The performance of the system is considered in Section \ref{sec:symbol_error_probability}. Simulation results are given in Section \ref{sec:sim}. Section \ref{sec:con} concludes the paper.

\emph{Notation}: $()^*$ and $|\cdot|$ denote conjugate and absolute values, respectively. $\mathcal{CN}(0,\sigma^2)$ and $\chi_2^2$ stand for complex Gaussian distribution with mean zero and variance $\sigma^2$ and chi-squared distribution with two degrees of freedom, respectively. $\mathrm{E}\{\cdot\}$ and $\mathrm{Var}\{\cdot\}$ denote expectation and variance operations, respectively. Both ${\mathrm{e}}^{(\cdot)}$ and $\exp(\cdot)$ show the exponential function. $\mathrm{Re}\{\cdot\}$ stands for the real part of a complex number.


\section{System Model}

\label{sec:system}
The wireless relay model under consideration has one source, one relay and one destination. The source communicates with the destination both directly and via the relay. Each node has a single antenna, and the communication between nodes is half duplex (i.e., each node is able to only send or receive in any given time). The channels from the source to the destination (SD), from the source to the relay (SR) and from the relay to the destination (RD) are denoted by $h_{\mathrm{sd}}[k]$, $h_{\mathrm{sr}}[k]$ and $h_{\mathrm{rd}}[k]$, respectively and they are assumed to be Rayleigh flat-fading. The channels are spatially uncorrelated and change continuously with Jakes' fading model \cite{microwave-jake}:
$
\mathcal{E} \{h_{\mathrm{ij}}[k]h_{\mathrm{ij}}^*[k+n]\}=J_0(2\pi f_{\mathrm{ij}} n),
$
where $J_0(\cdot)$ is the zeroth-order Bessel function of the first kind, $f_{\mathrm{ij}}$ is the maximum normalized Doppler frequency of the channel and $\mathrm{ij} \in \{\mathrm{sd,sr,rd}\}$ .

The $\log_2M$ information bits at time $k$ are transformed to $M$-PSK symbols as $v_m\in \mathcal{V}$ where $\mathcal{V}=\{{\mathrm{e}}^{j2\pi m/M},\; m=0,\dots, M-1\}$. Before transmission, the symbols are encoded differentially as
$ 
\label{eq:s-source}
s[k]=v_m\; s[k-1],\quad s[0]=1.
$

The transmission process is divided into two phases. Technically, either symbol-by-symbol or block-by-block dual-phase transmission protocol can be considered. In symbol-by-symbol transmission, first the source sends one symbol to the relay and the destination. Then, the relay re-transmits the symbol to the destination. Hence, two channel uses are two symbols apart. However, this causes frequent switching between reception and transmission which is not practical. Instead, in block-by-block transmission, a block of data is broadcasted in each phase and then two channel uses are one symbol apart. Although block-by-block transmission is considered in this paper, the analysis is the same for both cases as only the channel auto-correlation value is different.

In phase I,  symbol $s[k]$ is transmitted by the source to the relay and the destination. Let $P_0$ be the average source power. The received signal at the destination and the relay are
\begin{equation}
\label{eq:source_destination_rx}
y_{\mathrm{sd}}[k]=\sqrt{P_0}h_{\mathrm{sd}}[k]s[k]+w_{\mathrm{sd}}[k]
\end{equation}
and
\begin{equation}
\label{eq:relay_rx}
y_{\mathrm{sr}}[k]=\sqrt{P_0}h_{\mathrm{sr}}[k]s[k]+w_{\mathrm{sr}}[k]
\end{equation}
where $w_{\mathrm{sd}}[k],w_{\mathrm{sr}}[k]\sim \mathcal{CN}(0,1)$ are noise components at the destination and the relay, respectively.

The received signal at the relay is then multiplied by an amplification factor $A$, and re-transmitted to the destination. The corresponding received signal at the destination is
\begin{equation}
\label{eq:dest-rx1}
y_{\mathrm{rd}}[k]=Ah_{\mathrm{rd}}[k]y_{\mathrm{sr}}[k]+w_{\mathrm{rd}}[k],
\end{equation}
where $w_{\mathrm{rd}}[k]\sim \mathcal{CN}(0,1)$ is the noise at the destination. Substituting (\ref{eq:relay_rx}) into (\ref{eq:dest-rx1}) yields
\begin{equation}
\label{eq:destination-rx}
y_{\mathrm{rd}}[k]= A \sqrt{P_0}h[k]s[k]+w[k],
\end{equation}
where $h[k]=h_{\mathrm{sr}}[k]h_{\mathrm{rd}}[k]$ is the equivalent double-Rayleigh channel with zero mean and variance one \cite{SPAF-P, DGC-M} and
$
w[k]=A h_{\mathrm{rd}}[k]w_{\mathrm{sr}}[k]+w_{\mathrm{rd}}[k]
$
is the equivalent noise. It should be noted that for a given $h_{\mathrm{rd}}[k]$, $w[k]$ and $y_{\mathrm{rd}}[k]$ are complex Gaussian random variables with mean zero and variance $\sigma^2=A^2 |h_{\mathrm{rd}}[k]|^2+1$ and $\sigma^2(\rho+1)$, respectively, where $\rho$ is the average received SNR conditioned on $h_{\mathrm{rd}}[k]$, defined as
$
\label{eq:average-snr}
\rho=\frac{A^2 P_0 |h_{\mathrm{rd}}[k]|^2}{\sigma^2}.
$

The following section considers the differential detection of the combined received signals at the destination and evaluates its performance.



\section{Channel Model and Differential Detection}
\label{sec:ch-model}

The CDD was developed under the assumption that two consecutive channel uses are approximately equal. However, such an assumption is not valid for fast time-varying channels. To find the performance of differential detection in time-varying channels, we need to model both the direct and the cascaded channels with time-series models.

First, each individual link is modelled with an AR(1) model as follows \cite{AR1-ch}:
\begin{equation}
\label{eq:AR1-model}
h_{\mathrm{ij}}[k]=\alpha_{\mathrm{ij}} h_{\mathrm{ij}}[k-1]+\sqrt{1-\alpha_{\mathrm{ij}}^2}e_{\mathrm{ij}}[k]
\end{equation}
where $\alpha_{\mathrm{ij}}=J_0(2\pi f_{\mathrm{ij}}n)\leq 1$ is the auto-correlation of $h_{\mathrm{ij}}[k]$ and $e_{\mathrm{ij}}[k]\sim \mathcal{CN}(0,1)$ is independent of $h_{\mathrm{ij}}[k-1]$ for $ij\in\{\mathrm{sr},\mathrm{rd},\mathrm{sd}\}$. Also, $n=1$ for block-by-block transmission and $n=2$ for symbol-by-symbol transmission.

For the cascaded channel, multiplying $h_{\mathrm{sr}}[k]$ and $h_{\mathrm{rd}}[k]$ gives
\begin{equation}
\label{eq:hsr_hrd}
h[k]=\alpha h[k-1]+\Delta[k],
\end{equation}
where $\alpha=\alpha_{\mathrm{sr}}\alpha_{\mathrm{rd}}\leq 1$ is the equivalent auto-correlation and
\begin{equation}
\label{eq:Delta}
\begin{split}
\Delta[k]  &=\alpha_{\mathrm{sr}} \sqrt{1-\alpha_{\mathrm{rd}}^2} h_{\mathrm{sr}}[k-1] e_{\mathrm{rd}}[k]+\alpha_{\mathrm{rd}} \sqrt{1-\alpha_{\mathrm{sr}}^2}\\
&h_{\mathrm{rd}}[k-1] e_{\mathrm{sr}}[k]
+\sqrt{(1-\alpha_{\mathrm{sr}}^2)(1-\alpha_{\mathrm{rd}}^2)}e_{\mathrm{sr}}[k]e_{\mathrm{rd}}[k]
\end{split}
\end{equation}
represents the time-varying part of the equivalent channel. This part is a combination of three uncorrelated complex double Gaussian random variables \cite{DGC-M} and uncorrelated to $h[k-1]$. Since $\Delta[k]$ has a zero mean, its autocorrelation function is given as
\begin{equation}
\mbox{E}\{\Delta[k]  \Delta^*[k+n]\}=
\begin{cases}
1-\alpha^2 & \text{if} \; n=0, \\
 0 & \text{if} \; n \neq 0.
\end{cases}
\end{equation}
Therefore $\Delta[k]$ is a white noise process with the power of $\mbox{E}\{\Delta[k]  \Delta^*[k]\}=1-\alpha^2$.

Since the expression of $\Delta[k]$ as in \eqref{eq:Delta} is not easy to work with in terms of performance analysis, $\Delta[k]$ is approximated with one of its terms as
\begin{equation}
\label{eq:delta_h_hat}
\hat{\Delta}[k]= \sqrt{1-\alpha^2} {h}_{\mathrm{rd}}[k-1] {e}_{\mathrm{sr}}[k]
\end{equation}
The above approximation is also a white noise process with first and second order statistical properties identical to that of $\Delta[k]$ and it is also uncorrelated to $h[k-1]$.

By substituting \eqref{eq:delta_h_hat} into \eqref{eq:hsr_hrd}, the time-series model of the equivalent channel is given as
\begin{equation}
\label{eq:AR2-model}
h[k]=\alpha h[k-1]+\sqrt{1-\alpha^2} h_{\mathrm{rd}}[k-1]{e}_{\mathrm{sr}}[k]
\end{equation}

To validate the above model for the cascaded channel, its statistical properties are verified with the theoretical counterparts. Theoretical mean and variance of $h[k]$ are shown to be equal to zero and one, respectively, in \cite{SPAF-P} and \cite{DGC-M}. This can be seen by taking expectation and variance operations over \eqref{eq:AR2-model} to see that $\mbox{E}\{h[k]\}=0$ and $\mbox{Var}\{h[k]\}=1$. Also, the theoretical auto-correlation of $h[k]$ is obtained as the product of the auto-correlations of SR and RD channels in \cite{SPAF-P}. By multiplying both sides of \eqref{eq:AR2-model} with $h^*[k-1]$ and taking expectation and using the fact that $\mbox{E}\{\hat{\Delta}[k]h^*[k-1] \}=0$, it can be shown that
$
\label{eq:h[k]}
\mbox{E}\{h[k]h^*[k-1]\}=\alpha=\alpha_{\mathrm{sr}} \alpha_{\mathrm{rd}}.
$

\begin{figure*}[htb!]
\psfrag {h} [] [] [1] {$|h[k]|$}
\psfrag {delta} [] [] [1] {\eqref{eq:Delta}}
\psfrag {mdelta} [] [] [1] {$|\Delta[k]|$ or $|\hat{\Delta}[k]|$}
\psfrag {deltahat} [] [] [1] {\eqref{eq:delta_h_hat}}
\psfrag {PDF} [] [] [1] {pdf}
\psfrag {exact} [] [c] [1] {\eqref{eq:hsr_hrd}}
\psfrag {approx} [] [c] [1] {\eqref{eq:AR2-model}}
\psfrag {theory} [] [c] [1] {\eqref{eq:pdf-envelope}}
\centerline{\epsfig{figure={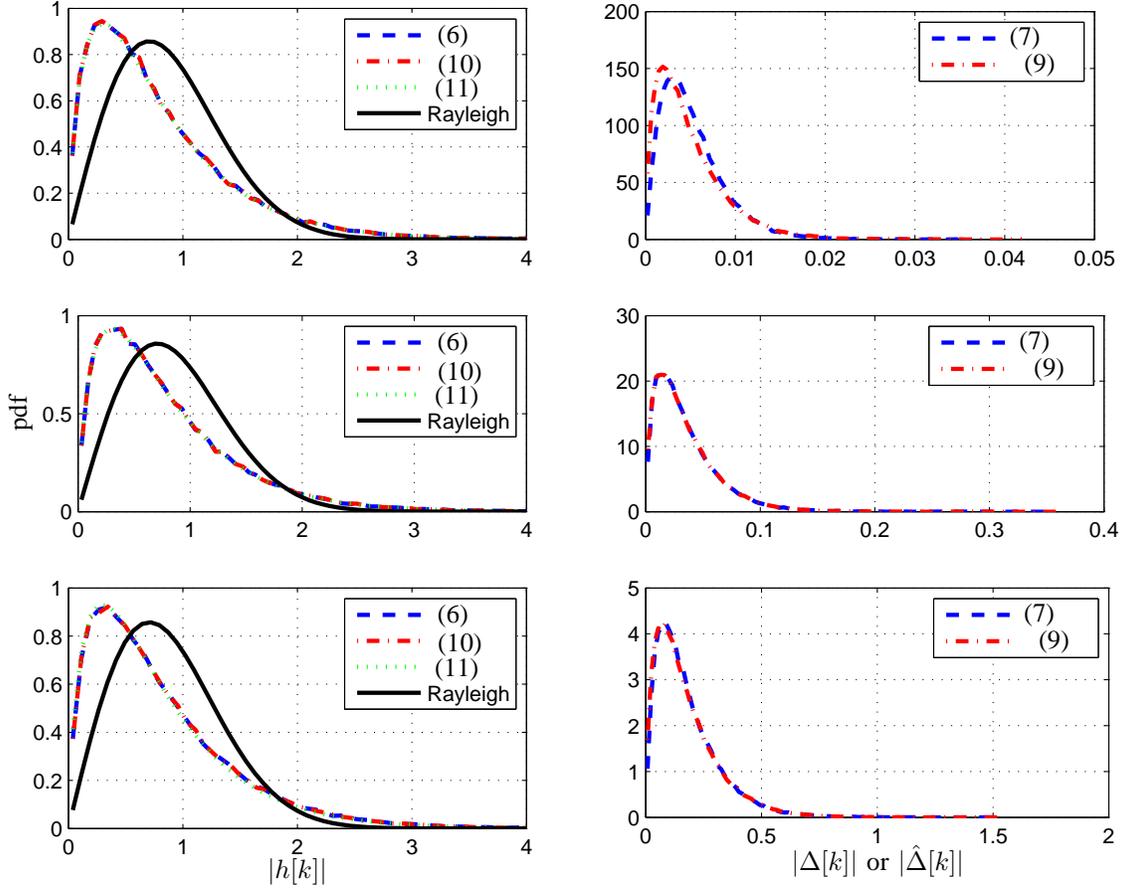},width=15cm}}
\caption{Theoretical pdf of $|h[k]|$ and obtained distributions of $|\Delta[k]|$, $|\hat{\Delta}[k]|$ and $|h[k]|$  in Scenario I (upper), Scenario II (middle) and Scenario III (lower), the scenarios are listed in Table I.}
\label{fig:pdf}
\end{figure*}

In addition, the theoretical pdf of the envelope $\lambda=|h[k]|$ is
\begin{equation}
\label{eq:pdf-envelope}
f_{\lambda}(\lambda)=4\lambda K_0\left( 2 \sqrt{\lambda^2}\right)
\end{equation}
where $K_0(\cdot)$ is the zeroth order modified Bessel function of the second kind \cite{SPAF-P}, \cite{DGC-M}.
To verify this, using Monte-Carlo simulation the histograms of $|h[k]|$, $|\Delta[k]|$ and $|\hat{\Delta}[k]|$, for different values of $\alpha$, are obtained for both models in \eqref{eq:hsr_hrd} and \eqref{eq:AR2-model}. The values of $\alpha$ are computed from the normalized Doppler frequencies given in Table \ref{table:scenarios} (which, as shown in Section \ref{sec:sim}, cover a variety of practical situations). These histograms along with the theoretical pdf of $|h[k]|$ are illustrated in Fig.~\ref{fig:pdf}. Although, theoretically, the distributions of $\Delta[k]$ and $\hat{\Delta}[k]$ are not exactly the same, we see that for practical values of $\alpha$ they are very close. Moreover, the resultant distributions of $h[k]$, regardless of $\Delta[k]$ or $\hat{\Delta}[k]$, are similar and follow the theoretical distribution. The Rayleigh pdf is depicted in the figure to illustrate the difference between the distributions of individual and the cascaded channels.


By substituting (\ref{eq:AR1-model}) and (\ref{eq:AR2-model}) into (\ref{eq:source_destination_rx}) and (\ref{eq:destination-rx}), respectively, one has
\begin{equation}
\label{eq:cddfast-source-destination}
y_{\mathrm{sd}}[k]=\alpha_{\mathrm{sd}} v_m\; y_{\mathrm{sd}}[k-1]+n_{\mathrm{sd}}[k],\\
\end{equation}
\begin{equation}
\label{eq:equivalent noise source-destination}
\begin{split}
n_{\mathrm{sd}}[k]&=w_{\mathrm{sd}}[k]- \alpha_{\sd} v_m\; w_{\mathrm{sd}}[k-1]\\
&+ \sqrt{1-\alpha_{\mathrm{sd}}^2} \sqrt{P_0} s[k]e_{\mathrm{sd}}[k],
\end{split}
\end{equation}
\begin{equation}
\label{eq:cddfast-relay-destination}
y_{\mathrm{rd}}[k]=\alpha v_m\; y_{\mathrm{rd}}[k-1]+n_{\mathrm{rd}}[k],
\end{equation}
\begin{equation}
\label{eq:equivalent noise relay-destination}
\begin{split}
n_{\mathrm{rd}}[k]&=w[k]- \alpha v_m\; w[k-1]\\
&+ \sqrt{1-\alpha^2}A\sqrt{P_0}h_{\mathrm{rd}}[k-1]s[k]{e}_{\mathrm{sr}}[k].
\end{split}
\end{equation}

Note that, for a given $h_{\mathrm{rd}}[k]$, the equivalent noise components $n_{\mathrm{sd}}[k]$ and $n_{\mathrm{rd}}[k]$ are  combinations of complex Gaussian random variables, and hence they are also complex Gaussian with variances
\begin{gather}
\label{eq:variance_eq_nosie1}
\sigma_{n_{\mathrm{sd}}}^2=1+\alpha_{\mathrm{sd}}^2+(1-\alpha_{\mathrm{sd}}^2)P_0\\
\label{eq:variance_eq_nosie2}
\sigma_{n_{\mathrm{rd}}}^2=\sigma^2 \left(1+\alpha^2+(1-\alpha^2)\rho \right)
\end{gather}


To achieve the \emph{cooperative} diversity, the two received signals from the two transmission phases are combined as
\begin{equation}
\label{eq:y_combined}
\zeta=b_0 y_{\mathrm{sd}}^*[k-1]y_{\mathrm{sd}}[k]+b_1 y_{\mathrm{rd}}^*[k-1]y_{\mathrm{rd}}[k]
\end{equation}
where $b_0$ and $b_1$ are combining weights. Using MRC technique \cite{Linear-Diversity}, the optimum combining weights, which takes into account the noise variance of each link, would be
$
b_0^{\mathrm{opt}}=\frac{\alpha_{\mathrm{sd}}}{\sigma_{n_\mathrm{sd}}^2}
$
and
$
b_1^{\mathrm{opt}}=\frac{\alpha}{\sigma_{n_\mathrm{rd}}^2}.
$
However, as can be seen from \eqref{eq:variance_eq_nosie2}, even for slow-fading channels with $\alpha=1$, the noise variance depends on the channel coefficients $h_{\mathrm{rd}}[k]$, which is not known in the system under consideration. To overcome this problem, for slow-fading channels, the average values of the noise variances, $E\{\sigma_{n_\mathrm{sd}}^2\}=2$ and $E\{\sigma_{n_\mathrm{rd}}^2\}=2(1+A^2)$ were utilized in \cite{DAF-Liu,DAF-DDF-QZ,General-DAF} to define the weights as
$
\label{eq:b0b1_cdd}
b_0^{\mathrm{cdd}}=\frac{1}{2}
$
and
$
b_1^{\mathrm{cdd}}=\frac{1}{2(1+A^2)}.
$
It is also shown in \cite{DAF-Liu,DAF-DDF-QZ,General-DAF} that these weights give a performance close to the performance of optimum combining.

For time-varying channels, the average noise variances of the two links (direct and cascaded) are $E\{\sigma_{n_\mathrm{sd}}^2\}=1+\alpha_{\mathrm{sd}}^2+(1-\alpha_{\mathrm{sd}}^2)P_0$ and $E\{\sigma_{n_\mathrm{rd}}^2\}=(1+\alpha^2)(1+A^2)+(1-\alpha^2)A^2P_0$, respectively.
Therefore, the new combining weights are proposed as follows:
\begin{equation}
\label{eq:b0_b1}
\begin{split}
&b_0=\frac{\alpha_{\mathrm{sd}}}{1+\alpha_{\mathrm{sd}}^2+(1-\alpha_{\mathrm{sd}}^2)P_0}\\
&b_1=\frac{\alpha}{(1+\alpha^2)(1+A^2)+(1-\alpha^2)A^2P_0}
\end{split}
\end{equation}

It can be seen that for slow-fading, $\alpha_{\mathrm{sd}}=\alpha=1$, which gives $b_0=b_0^{\mathrm{cdd}}$ and $b_1=b_1^{\mathrm{cdd}}$ as expected. However, for fast-fading channels, the weights change with the channel auto-correlation and source power. In essence, the new weights provide a dynamic combining of the received signals based on the fade rate of each link. The faster the channel changes in a communication link, the smaller portion of the received signal in that link is taken into account for detection.

Finally, the well known minimum Euclidean distance (ED) detection is expressed as \cite{Dig-comm-porakis},
$
\label{eq:ml-detection}
\hat{v}= \arg \min \limits_{v_m\in \mathcal{V}} |\zeta-v_m|^2.
$

The next section analyzes the performance of this detector.

\section{Error Performance Analysis}
\label{sec:symbol_error_probability}
Although, the practical combining weights given in \eqref{eq:b0_b1} are used in the detection process, finding the performance of the system with these weights appears very complicated (if not impossible). Instead, performance of the TVD system based on the optimum combining weights is carried out and used as a benchmark for the performance of the TVD approach.

Assume that symbol $v_1$ is transmitted and it is decoded erroneously as $v_2$, the nearest neighbour symbol, by the decoder. The corresponding PEP is defined as $
\label{eq:PEij}
P_s(E_{12})=P_s(v_1\rightarrow v_2).
$
An error occurs when
$
\label{eq:eulidian-distance}
|\zeta-v_1 |^2>|\zeta-v_2|^2
$
which can be simplified to
\begin{equation}
\label{eq:pep-cond1}
\text{Re} \left\lbrace (v_1-v_2)^*\zeta  \right\rbrace < 0.
\end{equation}

By substituting $\zeta$ from \eqref{eq:y_combined} and using $b_0=b_0^{\mathrm{opt}}$ and $b_1=b_1^{\mathrm{opt}}$ into the above inequality, the error event can be further simplified as $z>a$, where
\begin{equation}
\label{eq:z_gr_a}
\begin{split}
a&= |d_{\mathrm{min}}|^2 (\alpha_{\mathrm{sd}} b_0^{\mathrm{opt}} |y_{\mathrm{sd}}[k-1]|^2+\alpha b_1^{\mathrm{opt}} |y_{\mathrm{rd}}[k-1]|^2) \\
z&=-2 \text{Re} \left\lbrace d_{\mathrm{min}}^* ( b_0^{\mathrm{opt}} y_{\mathrm{sd}}^*[k-1] n_{\mathrm{sd}}[k] \right. \\  & \left. \hspace{1.8 in} + b_1^{\mathrm{opt}} y_{\mathrm{rd}}^*[k-1] n_{\mathrm{rd}}[k]) \right\rbrace
\end{split}
\end{equation}
and $d_{\mathrm{min}}=v_1-v_2$. Since $n_{\mathrm{sd}}[k]$ is Gaussian and also conditioned on $h_{\mathrm{rd}}[k]$, $n_{\mathrm{rd}}[k]$ is Gaussian, the variable $z$ conditioned on $y_{\mathrm{sd}}[k-1]$, $y_{\mathrm{rd}}[k-1]$ and $h_{\mathrm{rd}}[k]$ is Gaussian as well. Its mean, $\mu_z$, and variance, $\sigma^2_z$, conditioned on the above variables, are given as
\begin{equation*}
\label{eq:mean-z}
\mu_z=|d_{\mathrm{min}}|^2 \left( \frac{\alpha_{\mathrm{sd}}b_0^{\mathtt{opt}}}{P_0+1} |y_{\mathrm{sd}}[k-1]|^2+\frac{\alpha b_1^{\mathrm{opt}}}{\rho+1} |y_{\mathrm{rd}}[k-1]|^2\right)
\end{equation*}
\begin{equation*}
\label{eq:var-z}
\sigma_z^2=2|d_{\mathrm{min}}|^2\left( \alpha_{\mathrm{sd}} b_0^{\mathrm{opt}} |y_{\mathrm{sd}}[k-1]|^2
\right. \\\left. + \alpha b_1^{\mathrm{opt}} |y_{\mathrm{rd}}[k-1]|^2 \right)\nonumber
\end{equation*}

It follows that the conditional PEP can be written as
\begin{multline}
\label{eq:PEP_given_y_h }
P_s(E_{12}|y_{\mathrm{sd}},y_{\mathrm{rd}},h_{\mathrm{rd}})=
\text{Pr}(z> a|y_{\mathrm{sd}},y_{\mathrm{rd}},h_{\mathrm{rd}})\\
=Q\left( \frac{a-\mu_z}{\sigma_z}\right)=
Q\left( \sqrt{\Gamma_{\mathrm{sd}}+\Gamma_{\mathrm{rd}}}\right)
\end{multline}
where 
$\label{eq:Gamma-sd}
\Gamma_{\mathrm{sd}}= \frac{\gamma_{\mathrm{sd}} |d_{\mathrm{min}}|^2}{P_0+1} |y_{\mathrm{sd}}[k-1]|^2$, 
$
\label{eq:Gamma-rd}
\Gamma_{\mathrm{rd}}= \frac{\gamma_{\mathrm{rd}} |d_{\mathrm{min}}|^2}{\sigma^2 (\rho+1)} |y_{\mathrm{rd}}[k-1]|^2
$
and
$\gamma_{\mathrm{sd}}$ and $\gamma_{\mathrm{rd}}$ are defined as
\begin{gather}
\label{eq:gamma_source_destination}
\gamma_{\mathrm{sd}}=\frac{\alpha_{\mathrm{sd}}^2 P_0}{2P_0(1-\alpha_{\mathrm{sd}}^2)+4+\frac{2}{P_0}}\\
\label{eq:gamma_relay_destination}
\gamma_{\mathrm{rd}}=\frac{\alpha^2 \rho}{2\rho(1-\alpha^2)+4+\frac{2}{\rho}}
\end{gather}

The next step is to take the average of $P_s(E_{12}|y_{\mathrm{sd}},y_{\mathrm{rd}},h_{\mathrm{rd}})$ over the distribution of $|y_{\mathrm{sd}}[k-1]|^2$ and $|y_{\mathrm{rd}}[k-1]|^2$. This is done as follows:
\begin{multline}
\label{eq:PEP-h2-MGF}
P_s(E_{12}|h_{\mathrm{rd}}[k])=\\
\frac{1}{\pi} \int \limits_0^{\pi/2} M_{\Gamma_{\mathrm{sd}}} \left( - \frac{1}{2 \sin^2 \theta} \right) M_{\Gamma_{\mathrm{rd}}} \left( - \frac{1}{2 \sin^2 \theta} \right) \dd\theta
\end{multline}
where $M_{\Gamma_{\mathrm{sd}}}(\cdot)$ and $M_{\Gamma_{\mathrm{rd}}}(\cdot)$ are the moment-generating functions (MGFs) of
$\Gamma_{\mathrm{sd}}$ and $\Gamma_{\mathrm{rd}}$, respectively. Since $y_{\mathrm{sd}}[k-1]$ and also $y_\mathrm{rd}[k-1]$ conditioned on $h_{\mathrm{rd}}[k]$, are $\mathcal{CN}(0,P_0+1)$ and $\mathcal{CN}(0,\sigma^2(\rho+1))$, respectively, one has $|y_{\mathrm{sd}}[k-1]|^2\sim \frac{P_0+1}{2} \chi_2^2$ and $|y_{\mathrm{rd}}[k-1]|^2\sim \frac{\sigma^2(\rho+1)}{2} \chi_2^2$, respectively. Hence, the MGF of $\Gamma_{\mathrm{sd}}$ and $\Gamma_{\mathrm{rd}}$ can be shown to be \cite{probab-Miller}
$M_{\Gamma_{\mathrm{sd}}}(s)=\frac{1}{1-s \gamma_{\mathrm{sd}} |d_{\mathrm{min}}|^2}$ and $M_{\Gamma_{\mathrm{rd}}}(s)=\frac{1}{1-s \gamma_{\mathrm{rd}} |d_{\mathrm{min}}|^2}$.

By substituting the above MGFs into (\ref{eq:PEP-h2-MGF}), one obtains
\begin{multline}
\label{eq:PEP-h2-MGF2}
P_s(E_{12}|h_{\mathrm{rd}})=\\
\frac{1}{\pi} \int \limits_0^{\pi/2} \frac{1}{1+\frac{1}{2\sin^2 \theta}\gamma_{\mathrm{sd}} |d_{\mathrm{min}}|^2} \frac{1}{1+\frac{1}{2\sin^2 \theta}\gamma_{\mathrm{rd}} |d_{\mathrm{min}}|^2} \dd\theta
\end{multline}

The above integral can be solved by partial fraction technique and then averaged over the distribution of $|h_{\mathrm{rd}}[k]|^2$. However, this leads to a complicated expression without much insight. Instead, take the average over the distribution of $|h_{\mathrm{rd}}[k]|^2$, $f(\eta)=\exp(-\eta),\; \eta>0$ and the unconditioned PEP is given as
\begin{equation}
\label{eq:PEP-integral}
P_s(E_{12})=\\
\frac{1}{\pi} \int \limits_0^{\pi/2} \frac{I_1(\theta)}{1+\frac{1}{2\sin^2 \theta}\gamma_{\mathrm{sd}} |d_{\mathrm{min}}|^2}  \dd\theta
\end{equation}
\begin{equation}
\label{eq:I1}
\begin{split}
\text{where\quad} I_1(\theta)&=\int \limits_0^{\infty} \frac{e^{-\eta}}{1+\frac{1}{2\sin^2 \theta}\gamma_{\mathrm{rd}} |d_{\mathrm{min}}|^2} \dd\eta\\
&=\epsilon_1(\theta) \left[ 1+ (\beta_1-\beta_2(\theta)) e^{\beta_2(\theta)} E_1(\beta_2(\theta)) \right]
\end{split}
\end{equation}
\vspace*{-.5cm}
\begin{multline}
\label{eq: c1_beta1_beta2}
\epsilon_1(\theta)=\frac{4(1-\alpha^2)A^2P_0+8A^2}{\frac{1}{\sin^2(\theta)}\alpha^2A^2P_0|d_{\mathrm{min}}|^2+4(1-\alpha^2)A^2P_0+8A^2}\\
\beta_1=\frac{4}{2(1-\alpha^2)A^2P_0+4A^2}\\
\beta_2(\theta)=\frac{8}{\frac{1}{\sin^2(\theta)}\alpha^2A^2P_0|d_{\mathrm{min}}|^2+4(1-\alpha^2)A^2P_0+8A^2}
\end{multline}
and $E_1(x)=\int \limits_x^{\infty} \frac{{\mathrm{e}}^{-t}}{t}\dd t$ (exponential integral function).
The integral in \eqref{eq:PEP-integral} can be computed numerically to find the PEP.

It can be verified that, for DBPSK, the expression in \eqref{eq:PEP-integral} gives the exact bit-error rate (BER). On the other hand, for higher-order $M$-PSK constellations, the nearest-neighbour approximation \cite{Dig-comm-porakis} shall be applied to obtain the overall symbol-error rate (SER) as
$
P_s(E)\approx 2 P_s(E_{12}),
$
and the average BER for Gray mapping as
$
P_b(E)\approx \frac{2}{\log_2 M} P_s(E_{12}).
$

Finding an upper bound for the PEP expression can help to get more insights about the system performance. For $\theta=\frac{\pi}{2}$, \eqref{eq:PEP-integral} is bounded as
$
P_s(E_{12})\leq \frac{I_1(\frac{\pi}{2})}{2+\gamma_{\mathrm{sd}}|d_{\mathrm{min}}|^2}.
$
Based on the definition of $\gamma_{\mathrm{sd}}$ and $I_1(\frac{\pi}{2})$, in \eqref{eq:gamma_relay_destination} and \eqref{eq:I1}, it can be seen that the error probability depends on the fading rates, $\alpha_{\mathrm{sd}}$ and $\alpha$, of both the direct and cascaded channels. If all channels are very slow, $\alpha_{\mathrm{sd}}= 1$ and $\alpha= 1$ and it can be verified that $\gamma_{\sd} \propto P_0$ and $I_1(\frac{\pi}{2})\propto \frac{1}{P_0}$, thus the diversity of two is achieved. On the other hand, if because of the mobility of the nodes respect to each other, the channels become time-selective, the terms $(1-\alpha_{\mathrm{sd}}^2)P_0$ and $(1-\alpha^2)P_0$ in the denominator of $\gamma_{\mathrm{sd}}$ and $I_1(\frac{\pi}{2})$ become significant in high SNR. This leads to a degradation in the effective values of $\gamma_{\mathrm{sd}}$ and $\gamma_{\mathrm{rd}}$ and hence a severe degradation in the overall performance and the achieved diversity.

It is also informative to examine the expression of PEP at high SNR values. In this case, $\lim \limits_{P_0\rightarrow \infty} \gamma_{\mathrm{sd}}= \frac{\alpha_{\mathrm{sd}}^2}{2(1-\alpha_{\mathrm{sd}}^2)}$ and $\lim \limits_{P_0\rightarrow \infty} E[\gamma_{\mathrm{rd}}]= \frac{\alpha^2}{2(1-\alpha^2)}$, which is independent of $|h_{\mathrm{rd}}[k]|$. Therefore, by substituting the converged values into \eqref{eq:PEP-h2-MGF2}, one can see that the error floor appears. Specifically,
\begin{multline}
\label{eq:PEP-floor1}
\hspace*{-0.5cm}\lim \limits_{P_0 \rightarrow \infty} P_s(E_{12})= 
\frac{1}{2} -\frac{\alpha_{\mathrm{sd}}^2(1-\alpha^2)}{2(\alpha_{\mathrm{sd}}^2-\alpha^2)}\sqrt{\frac{\alpha_{\mathrm{sd}}^2|d_{\mathrm{min}}|^2}{\alpha_{\mathrm{sd}}^2|d_{\mathrm{min}}|^2+4(1-\alpha_{\mathrm{sd}}^2)}}\\
+\frac{\alpha^2(1-\alpha_{\mathrm{sd}}^2)}{2(\alpha_{\mathrm{sd}}^2-\alpha^2)}\sqrt{\frac{\alpha^2|d_{\mathrm{min}}|^2}{\alpha^2|d_{\mathrm{min}}|^2+4(1-\alpha^2)}},
\end{multline}
when $\alpha_{\mathrm{sd}} \neq \alpha, \;\; \alpha_{\mathrm{sd}}, \alpha<1$, and
\begin{multline}
\label{eq:PEP-floor2}
\lim \limits_{P_0 \rightarrow \infty} P_s(E_{12})= 
\frac{1}{2}
\left\lbrace
1- \sqrt{\frac{\alpha|d_{\mathrm{min}}|^2}{\alpha|d_{\mathrm{min}}|^2+4(1-\alpha^2)}} \right. \\
\left. \left( 1+ \frac{1-\alpha^2}{\alpha^2|d_{\mathrm{min}}|^2+4(1-\alpha^2)}\right)
\right\rbrace
\end{multline}
when $\alpha_{\mathrm{sd}}=\alpha <1.$

It should be noted that the PEP and the error floor expressions are obtained based on the optimum combining weights. Therefore, they give lower bounds for the actual performance curves (obtained by simulation) of the system when the proposed weights are used.

\vspace*{-0.15cm}
\section{Simulation Results}
\label{sec:sim}
In all simulations, the channels $h_{\mathrm{sd}}[k]$, $h_{\mathrm{sr}}[k]$ and $h_{\mathrm{rd}}[k]$ are generated individually according to the simulation method of \cite{ch-sim}. Based on the normalized Doppler frequencies of the channels, three different scenarios are considered: (I) all channels are slow-fading, (II) SD and SR channels are fairly fast and RD channel is slow-fading, (III) SD and SR channels are very fast and RD channel is fairly-fast fading. The normalized Doppler values of the three scenarios are shown in Table \ref{table:scenarios}. To acquire a sense about the speed of communication nodes in each scenario, a typical wireless system with carrier frequency $f_c=2$ GHz and symbol time $T_s=0.1$ ms is assumed. Then, the corresponding Doppler shifts would be around $f_D=10, 100, 500$ Hz which are related to the vehicle speeds of $v=\frac{c f_D}{f_c}=5, 54, 270$ km/hr, respectively, where $c=3\times 10^8$ m/s is the speed of light.\footnote{A typical vehicle speed of 75 km/hr is usually assumed in the literature.} It can be seen that the three scenarios selected for simulations cover a wide variety of practical situations. \vspace*{-0.35cm}
\begin{table}[!h]
\begin{center}
\caption{Three simulation scenarios.}
\label{table:scenarios}
  \begin{tabular}{ |c | c| c| c | }
    \hline
				& $f_{\mathrm{sd}}$ & $f_{\mathrm{sr}}$ & $f_{\mathrm{rd}}$  \\ \hline\hline
{Scenario I}    & .001              & .001              & .001  			 \\ \hline
{Scenario II}   & .01               & .01 			    & .001   			  \\ \hline
{Scenario III}  & .05               & .05 			    & .01   			  \\
    \hline
  \end{tabular}
\end{center}
\end{table}
\vspace*{-0.5cm}

\begin{figure}[t!]
\psfrag {P(dB)} [][b] [.8]{$P$ (dB)}
\psfrag {BER} [] [] [.8] {BER}
\psfrag {Theoritical Optimum} [] [c] [.6] {Theoretical Optimum}
\psfrag {Scenario I} [] [] [.8] {Scenario I}
\psfrag {Scenario II} [] [] [.8] {Scenario II}
\psfrag {Scenario III} [] [b] [.8] {Scenario III}
\centerline{\epsfig{figure={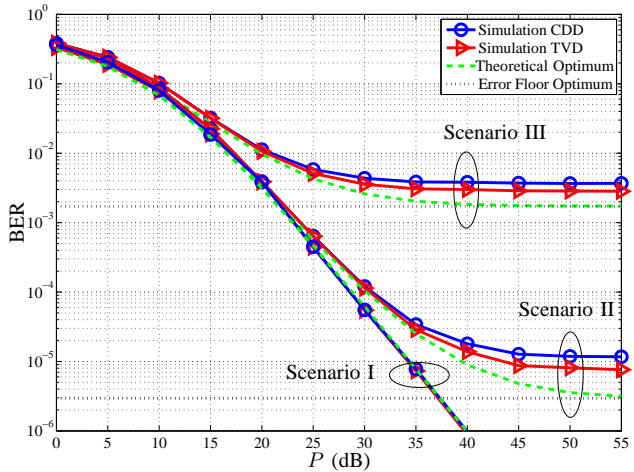},width=8.5cm}}
\caption{Theoretical and simulation BER of D-AF relaying for a three-node relay network in three scenarios using DQPSK.}
\label{fig:scs_m4}
\end{figure}

\begin{figure}[!ht]
\psfrag {P(dB)} [][b] [.8]{$P$ (dB)}
\psfrag {BER} [] [] [.8] {BER}
\psfrag {Scenario I} [] [] [.8] {Scenario I}
\psfrag {Scenario II} [] [] [.8] {Scenario II}
\psfrag {Scenario III} [] [b] [.8] {Scenario III}
\psfrag {Theoritical Optimum} [] [c] [.6] {Theoretical Optimum}
\centerline{\epsfig{figure={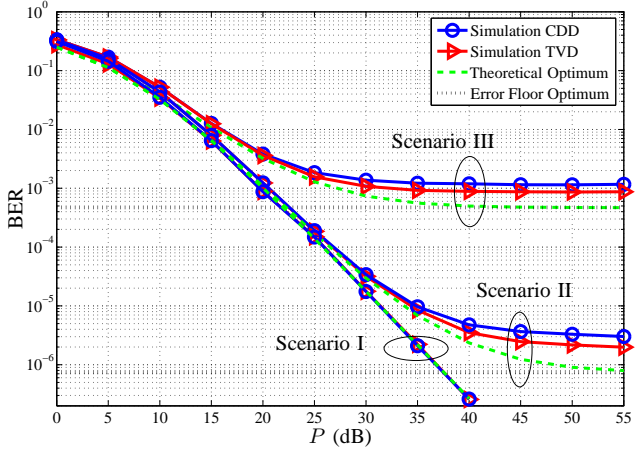},width=8.5cm}}
\caption{Theoretical and simulation BER of D-AF relaying for a three-node relay network in three scenarios using DBPSK.}
\label{fig:scs_m2}
\end{figure}

In each scenario, binary data is differentially encoded for $M=2,\;4$ constellations. Block-by-block transmission is utilized for all scenarios. The amplification factor at the relay is fixed to $A=\sqrt{\frac{P_1}{P_0+1}}$ to normalize the average relay power to $P_1$. Equal power allocation\footnote{Investigating the power allocation between the source and relay nodes is interesting  but beyond the scope of this paper.} between the source and the relay is used for a given $P$, i.e., $P_0=P_1=\frac{P}{2}$. 
The BER values obtained by simulation of CDD and TVD approaches are plotted versus $P$ in Fig.~\ref{fig:scs_m2} and Fig.~\ref{fig:scs_m4} with solid lines, for DBPSK, DQPSK, respectively. For computing the theoretical BER values, first the value of $\alpha$ and $\alpha_{\mathrm{sd}}$ are computed for each scenario. Also, $|d_{\min}|^2=4   \sin^2(\frac{\pi}{M})$ for $M$-PSK symbols are computed which gives $|d_{\min}|^2=4,\;2$ for $M=2,\;4$, respectively. Then, the corresponding theoretical BER values are plotted in the figures with dashed lines.

As can be seen in Fig.~\ref{fig:scs_m2}, in Scenario I, the diversity of two is achieved for both CDD and TVD approaches. Both error probabilities are monotonically decreasing with $P$ and are consistent with theoretical values. The simulation and theoretical values are tight in this case as all channels are slow-fading. In Scenario II, which involves two fairly fast-fading channels, the same results as the previous scenario can be seen for SNR less than 25 dB. But after this point, the plots gradually deviate from their slow-fading counterparts and reach an error floor at high $P$. The theoretical error floor is also calculated from (\ref{eq:PEP-floor1}) and plotted in the figure with dotted lines. This phenomena starts earlier, around 10 dB, in Scenario III, and the performance degradation is much more severe since all channels are fast. As expected, in both Scenarios II and III, the theoretical plot with the optimum gains is the best which, in fact, cannot be achieved in practice because no instantaneous CSI is available. However, this could serve as an useful performance lower bound for the system. On the other hand, the TVD system with the proposed weights outperforms the CDD system and performs closer to the optimum case. Similar behaviours can be seen in Fig.~\ref{fig:scs_m4} when using DQPSK. 

\balance

\vspace*{-0.15cm}

\section{Conclusion}
\label{sec:con}
The performance of differential amplify-and-forward relaying for a three-node relay network has been analyzed for time-varying channels. Based on the channel fade rates, new weights were proposed for combining the received signals at the destination. The obtained error probability expression shows that the error performance depends on the fading rates of the direct and the cascaded channels. For fast-fading channels, a high fading rate can lead to a severe degradation in the error probability. It was also shown that there exists an error floor at high SNR in time-varying channels and such an error floor was determined in terms of the channel auto-correlations. The analysis and formulation in this paper can be useful in designing efficient relay networks with high mobility and gives important insights in both theory and practice.

\vspace*{-0.15cm}

\section*{Acknowledgements}
The authors are thankful to Prof. Eric Salt and Dr. Tung Pham for valuable
discussions during the course of this work.

\bibliographystyle{IEEEbib}
\bibliography{references}

\end{document}